\documentclass[a4paper,11pt]{article}
\pdfoutput=1 

\usepackage{jheppub} 
\usepackage{multirow}
\usepackage[T1]{fontenc} 

\usepackage{lineno}

\title{\boldmath\huge{Potential of octant degeneracy resolution in JUNO} }

\author[a,1]{M.V. Smirnov,\note{Corresponding author.}}
\author[a]{Zh.J. Hu,}
\author[a]{S.J. Li}
\author[a,1]{{\rm and} J.J. Ling}

\affiliation[a]{Department of physics, Sun Yat-Sen University,\\Guangzhou 510275, China}

\emailAdd{gear8mike@gmail.com}
\emailAdd{lingjj5@mail.sysu.edu.cn}

\abstract
{
This research continues to focus on the idea using cyclotronic antineutrino source for purposes of neutrino physics.
Long baseline experiments suffer from degeneracies and correlations between $\Theta_{23}$, $\delta_{\rm CP}$ and the mass hierarchy. 
However the combination of a superconductive cyclotron and a big liquid scintillator detector like JUNO in a medium baseline experiment, which doesn't depend on the mass hierarchy, may allow us to determine whether the position of the mixing angle $\Theta_{23}$ is in the lower octant or the upper octant. Such an experiment would improve the precision of the $\Theta_{23}$ measurement to a degree which depends on the CP-phase. } 

\begin{document} 
\maketitle
\flushbottom

\section{The problem of octant degeneracy}
\label{sec:intro}

\par
\indent 

In the framework of 3-flavor neutrino mixing through Pontecorvo-Maki-Nakagawa-Sakata~\cite{PMNS} unitary mixing matrix:
\begin{equation}
\label{eq_1}
U_{\rm PMNS}=\left(\begin{array}{ccc}
U_{\rm e1}&U_{\rm e2}&U_{\rm e3} \\
U_{\rm \mu1}&U_{\rm \mu2}&U_{\rm \mu3} \\ 
U_{\rm \tau1}&U_{\rm \tau2}&U_{\rm \tau3} \\
\end{array}\right),
\end{equation}
in the standard parametrization $\sin^2(\Theta_{23})$ and $\cos^2(\Theta_{23})$ can be expressed as:
$$\sin^2(\Theta_{23})=\frac{|U_{\rm \mu3}|^2}{1-|U_{\rm e3}|^2},\qquad \cos^2(\Theta_{23})=\frac{|U_{\rm \tau3}|^2}{1-|U_{\rm e3}|^2}.$$
It is clear that if $\Theta_{23}=45^{\rm o}$, then mixing between $\nu_{\mu}$ and $\nu_{\tau}$ becomes maximal.
This would indicate symmetry between the $\nu_e\to\nu_{\mu}$ and 
$\nu_e\to\nu_{\tau}$ oscillation processes.
The octant problem refers to the degeneracy between $\Theta_{23}$ and $\pi/2 -\Theta_{23}$, because the mixing angle enters in the oscillation probability as a term within $\sin(2\Theta_{23})$.
However the degeneracy between the lower octant (LO) and the upper octant (UO) can be eliminated, if a measurement is sensitive to terms with $\sin(\Theta_{23})$ or $\cos(\Theta_{23})$.
Until recently, there was a quite a large uncertainty in the measurements of $\sin^2(\Theta_{23})$. $\sin^2(\Theta_{23})=0.35-0.65$ (90\%C.L.) for normal hierarchy (NH) and $\sin^2(\Theta_{23})=0.34-0.67$ (90\%C.L.) for inverted hierarchy (IH) from combined analysis of MINOS experiment~\cite{MINOS_2014}.
After that T2K reported the best fit value of $\sin^2(\Theta_{23})=0.532$(NH) and $\sin^2(\Theta_{23})=0.534$(IH) with smaller uncertainty and consistent with hypothesis of maximal mixing~\cite{T2K}. 
But recent data from the NO$\nu$A experiment now favors $\Theta_{23}$ in either LO or UO, and disfavors of maximal mixing at 0.8$\sigma$ significance~\cite{NOvA_2018}.   

Since the leading approximation of oscillation probability for reactor experiments does not depend on the mixing angle $\Theta_{23}$, the current scientific program of JUNO~\cite{juno} will not allow for a solution to the problem of octant degeneracy.
However precise measurements of ${\bar{\nu }}_{e}$ appearance from ${\bar{\nu }}_{\mu}$ disappearance would provide a good possibility for improvement of this issue. 

\section{Methodology of the numerical analysis}
\subsection{Proposal of the experimental setup}
\par
\indent 

The full description of our proposal is presented in~\cite{CPV_my}, which is based on the DAE$\delta$ALUS experiment project~\cite{daedalus}.
It is worthwhile to summarize the main components of the previous research. 
We suggested using the appearance channel for electron antineutrinos from muon antineutrinos.
In the framework of standard  three neutrino  mixing theory the oscillation probability can be expressed as~\cite{Bilenky}:
\begin{equation}
\begin{split}
\label{eq_21}
P\big({\bar{\nu }}_{\mu }{\to }{\bar{\nu }}_{e}\big)&=\sin ^{2}\theta _{23}\sin ^{2}2\theta _{13}\sin ^{2}\Delta _{31}+
\cos ^{2}\theta _{23}\sin ^{2}2\theta _{12}\sin ^{2}\Delta _{21}+\\
&+\sin 2\theta _{13}\sin 2\theta _{23}\sin 2\theta _{12}\sin \Delta _{31}\sin \Delta _{21}\cdot \cos (\Delta _{31}- \delta _{\mathrm{CP}}),
\end{split}
\end{equation}
where $\Delta_{ij}=\Delta m_{ij}^2\cdot L/(4E_{\nu})$; $\Delta m_{ij}^2$ -- the neutrino mass squared difference; $L$ -- the distance between source and detector; $E_{\nu}$ -- neutrino energy; $\delta _{\mathrm{CP}}$ -- Dirac phase of CP violation.
The source of  $\bar{\nu}_\mu$ is three-body decay of $\mu^+$ from decay at rest of the stopped $\pi^{+}$, which will be produced by a superconductive cyclotron~\cite{conrad}.
The contribution to antineutrino spectrum is around $10^{-4}$ from $\pi^{-}$, which are created together with $\pi^{+}$~\cite{daedalus}.
Two cyclotrons (near and far) will be located at distances 1.5 km and 20 km respectively. 
The power of the near cyclotron is 1 MW.
It is needed as a flux monitor.
There are two options for the power of the far cyclotron: 5 MW and 10 MW.
We are planning to use JUNO as a liquid scintillator detector, which has a total mass of 20 kt.  
The expected exposure time of the experiment is 10 years.
NH is assumed, because at the distance 20 km the experiment is insensitive to mass hierarchy.
\begin{figure}[ht]
\centering
\includegraphics[scale=0.47]{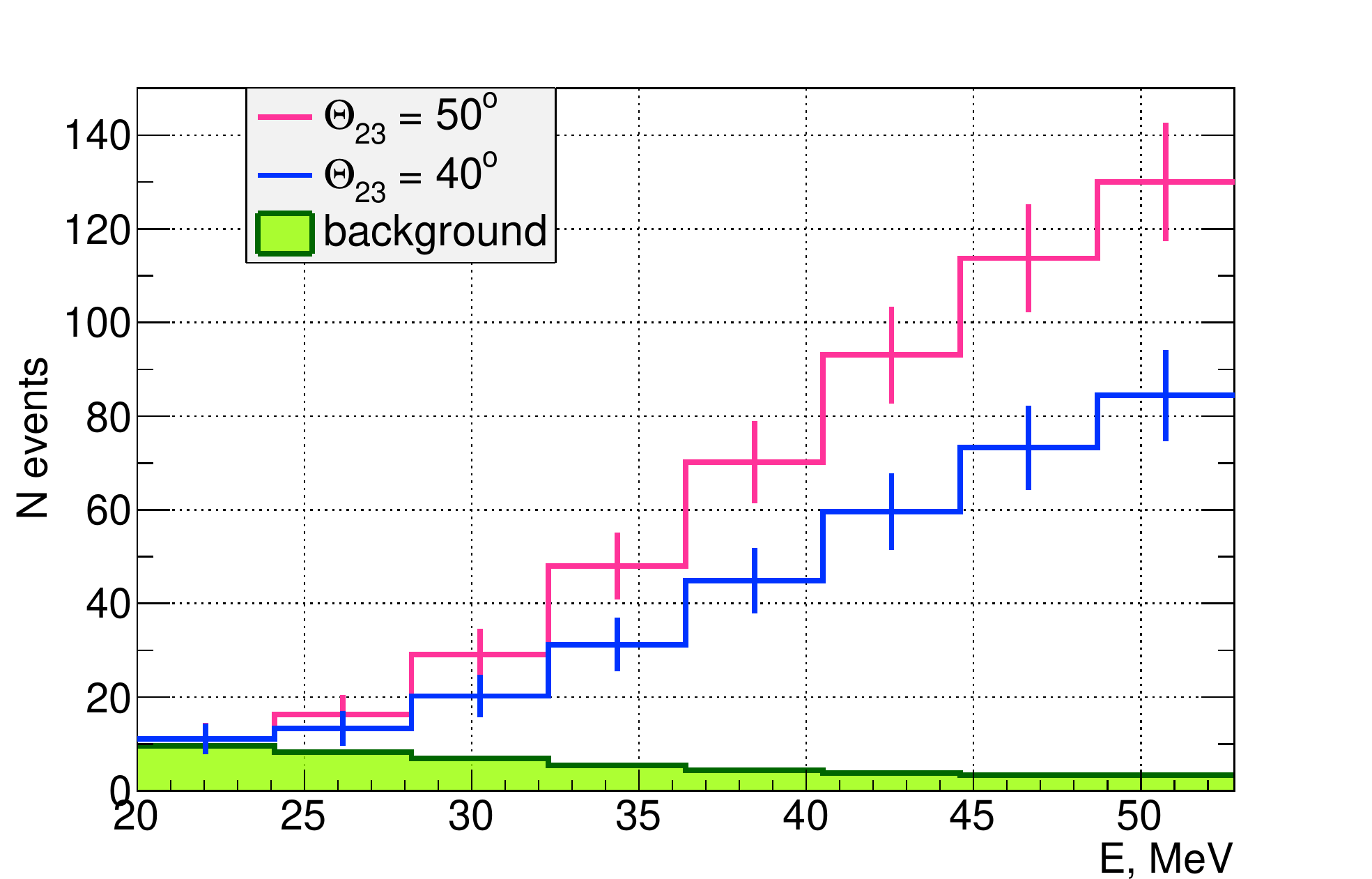}
\caption{\label{fig:0}The shape of the IBD-event spectrum as a function of energy for two different values of $\Theta_{23}$ (assumed 10 MW power of the far cyclotron, 200 kt$\cdot$year an exposure time, $\delta _{\mathrm{CP}}=-\pi/2$ ). The green area shows the background shape.}
\end{figure}
The estimated IBD-event spectrum as a function of energy is depicted in Figure~\ref{fig:0}. 
It is clear, that neutrino rate increases with the mixing angle $\Theta_{23}$.

\subsection{Statistical evaluation of MC simulations}
\par
\indent

Event rate analysis is based on statistical treatment of the expected IBD signal rate inside the detector.
Initial muon antineutrinos have a continuous spectrum with an endpoint of 52.8 MeV. 
In order to exclude a significant part of the atmospheric background, we chose the energy window between 20 and 52.8 MeV. 
However this is not sufficient to ignore the background completely. 

The current statistical analysis is split on two parts. The first part is the sensitivity to octant degeneracy. 
The second part is about the precise measurement of $\Theta_{23}$.

\subsubsection{\label{sec_3_3_1}Sensitivity to discovery of true octant}
Here we follow the so-called classical method of calculating a confidence level.
This method is based on the calculation of a $\Delta\chi^2$ function, which, as Wilks's theorem predicts~\cite{wilks}, should follow a chi-square distribution. 
The number of degrees of freedom can be calculated as the difference between degrees of freedom of initial chi-square functions. 
Usually, this number is equal to the quantity of estimating parameters.  
In our case, there is only one parameter -- $\Theta_{23}$.

A $\chi^2$ distribution with one degree of freedom has the same distribution as the square of a single normally distributed variable~\cite{num_stat}.
Therefore standard Gaussian confidence levels 1$\sigma$(68.3\%), 2$\sigma$(95.4\%), 3$\sigma$(99.7\%) etc. correspond to values of $\chi^2$: 1, 4, 9 and etc.

In general the sensitivity to octant degeneracy can be calculated through the minimization of a $\Delta\chi^2$ function, which is given by:
\begin{equation}
\label{eq_5}
\Delta\chi^2=|\chi_{\rm min}^2(90^{\rm o}-\Theta_{23})-\chi_{\rm min}^2(\Theta_{23})|,
\end{equation}
where ``min'' means, that both chi-square functions $\chi^2(90^{\rm o}-\Theta_{23})$ and $\chi^2(\Theta_{23})$ have to be minimized through their parameter space; $\Theta_{23}$ is a scanning parameter, which is fixed for each iteration of MC cycle.
In our case, the chi-square function has only one minimum, which is close to the test-true value of $\Theta_{23}$. 
In the opposite octant this function always increases. 
Consequently we need to redefine the $\Delta\chi^2$ function as:
\begin{equation}
\label{eq_5_1}
\Delta\chi^2=\chi_{\rm min}^2(45^{\rm o})-\chi_{\rm min}^2(\Theta_{23}),
\end{equation}
where $45^{\rm o}$ corresponds to a border between two octants.

We use the chi-square function presented in~\cite{Das_lbno,Huber}. 
\begin{equation}
\label{eq_6}
\chi^2(\Theta_{23})=\chi^2_{pull}+\chi^2_{prior},
\end{equation}
where {\it pull}-term includes Poisson statistics while taking into account the background and flux normalization. 
Additional Gaussian penalties are also added.
\begin{equation}
\label{eq_7}
\chi^2_{pull}=2\sum_{i=1}^{N_b}\Big{[}\mu_i-n_i+n_i\cdot\ln\frac{n_i}{\mu_i}\Big{]}+\frac{s^2}{\sigma^2_s}+\frac{b^2}{\sigma^2_b}.
\end{equation}
Here $N_b$ -- total number of bins in the histogram; $\mu_i$ -- predicted counts in the $i$-th bin; $n_i$ -- observed counts in the $i$-th bin;
$s$ and $b$ -- so-called nuisance parameters for signal and background respectively; $\sigma_s$ and $\sigma_b$ -- systematic errors for signal and background counts.
$\mu_i$ is represented by next equation:
$$\mu_i=N_s^i\cdot(1+s)+N_{bkg}^i\cdot(1+b),$$
where $N_s^i$ and $N_{bkg}^i$ -- number of counts in the $i$-th bin for signal and background respectively.
The second {\it prior}-term of the function~\eqref{eq_6} corresponds to uncertainties of oscillation parameters and can be written as:
 \begin{equation}
 \label{eq_8}
 \chi^2_{prior}=\sum_{j=1}^{N_p}\frac{(\eta_j-\eta_j^o)^2}{(\delta\eta_j)^2}, 
\end{equation}
where  $N_p$ -- quantity of oscillation parameters; $\eta_j$ -- $j$-th oscillation parameter; $\eta_j^o$ -- best fit value of $\eta_j$; $\delta\eta_j$ -- one sigma error of $\eta_j^o$.

\subsubsection{The accuracy of $\Theta_{23}$ measurement}
The estimation of the accuracy of measurement for the current best fit value of $\Theta_{23}$ can be obtained by minimizing the chi-square function~\eqref{eq_6} through the whole parameter space.
It should be emphasized, that from recent experimental data~\cite{NOvA_2018} the best fit value of $\Theta_{23}$ is split between LO and UO. And we also use two values of $\Theta_{23}$ in the calculation of precision. 

Further, we give a set of the oscillation parameters and their uncertainties taken from PDG in Table~\ref{tab_1}. 
\begin{table}[tbp]
\centering
\renewcommand{\arraystretch}{1.25}
\begin{tabular}{lccccc}
\hline
\hline
$\eta_j$ & $\Delta m_{21}^2\cdot10^{-5}~{\rm eV}^2$ & $\Delta m_{32}^2\cdot10^{-3}~{\rm eV}^2$ &$\sin^2(\Theta_{12})$ & $\sin^2(\Theta_{23})$ & $\sin^2(\Theta_{13})\cdot10^{-2}$ \\
\hline 
\hline
$\eta_j^o$ & 7.53 & 2.51 & 0.307 & $^{\rm0.597 (UO)}_{\rm0.417 (LO)}$ & 2.12\\
$\delta\eta_j$ & 0.18 & 0.05 & 0.013 & 0.026 & 0.08\\
\hline
\end{tabular}
\caption{\label{tab_1} The list of oscillation parameters and their uncertainties from PDG~\cite{PDG}. Most of them were used in {\it prior}-term of the chi-square function for our calculations, except the parameter of interest -- $\Theta_{23}$. The normal hierarchy is assumed.}
\end{table}

\subsubsection{Monte-Carlo simulations}
The expected electron antineutrino event spectra at a distance of 20 km were simulated using the Monte-Carlo method including oscillation.
The energy resolution of the JUNO detector is 3\% per 1 MeV.
The beam power of the far cyclotron is 5 or 10 MW with systematic flux uncertainty $\sigma_s=2$\%, which includes the uncertainties of shape and normalization.
We treat neutral current events (NC) as a background.
The initial estimation  gives 439 NC events for an exposure time of 200 kt$\cdot$year with a duty factor of 33\%. 
Using a technique from~\cite{Randolph} which is based on the signals coincidence and pulse shape discrimination, this background can be reduced significantly to 33 NC events. 
Adding also fast neutron and charge current atmospheric events, the total background will equal 45 events. 
The last number is used in simulations with systematic uncertainty $\sigma_b=5$\%.    

To investigate the sensitive region of octant degeneracy, 1K MC ``fake'' experiments were calculated for each sample with particular fixed values of $\delta_{\rm CP}$. 
We did not apply any constraints to the parameter $\Theta_{23}$.
Both  parts of $\Delta\chi^2$ in equation~\eqref{eq_5} were minimized using the ROOT package Minuit ~\cite{minuit,minuit2}.
Finally, the sensitivity region was calculated as defined in section~\ref{sec_3_3_1}. 

In order to evaluate the potential of JUNO to accurately measure the mixing angle $\Theta_{23}$, 5K MC ``fake'' experiments were simulated for each sample with particular fixed value of $\delta_{\rm CP}$. The chi-square function~\eqref{eq_6} was minimized through all parameter space.
Then a histogram was filled by the extracted values of $\Theta_{23}$. 
The shape of the histogram is Gaussian, because we supposed all uncertainties of parameters have Gaussian distribution.
The 1$\sigma$ error of $\Theta_{23}$ was obtained as a standard deviation of the aforementioned histogram.
This procedure has been repeated for the whole range of CP-phase from $-\pi$ to $\pi$. 

\section{Results} 
\par
\indent

\begin{figure}[ht]
\centering
\begin{minipage}[b]{0.44\linewidth}
\centering
\includegraphics[scale=0.33]{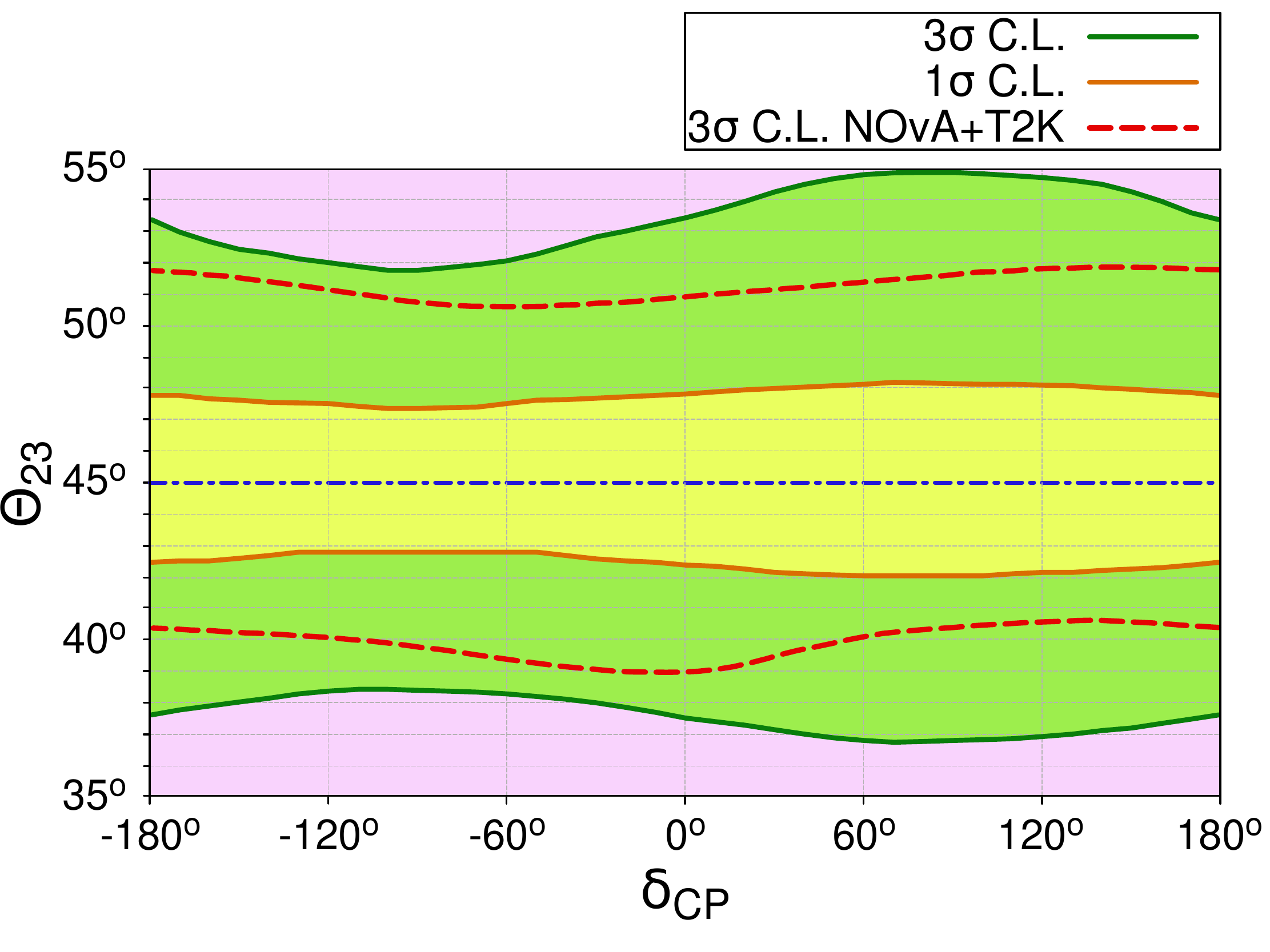}
\end{minipage}
\qquad
\begin{minipage}[b]{0.48\linewidth}
\centering
\includegraphics[scale=0.33]{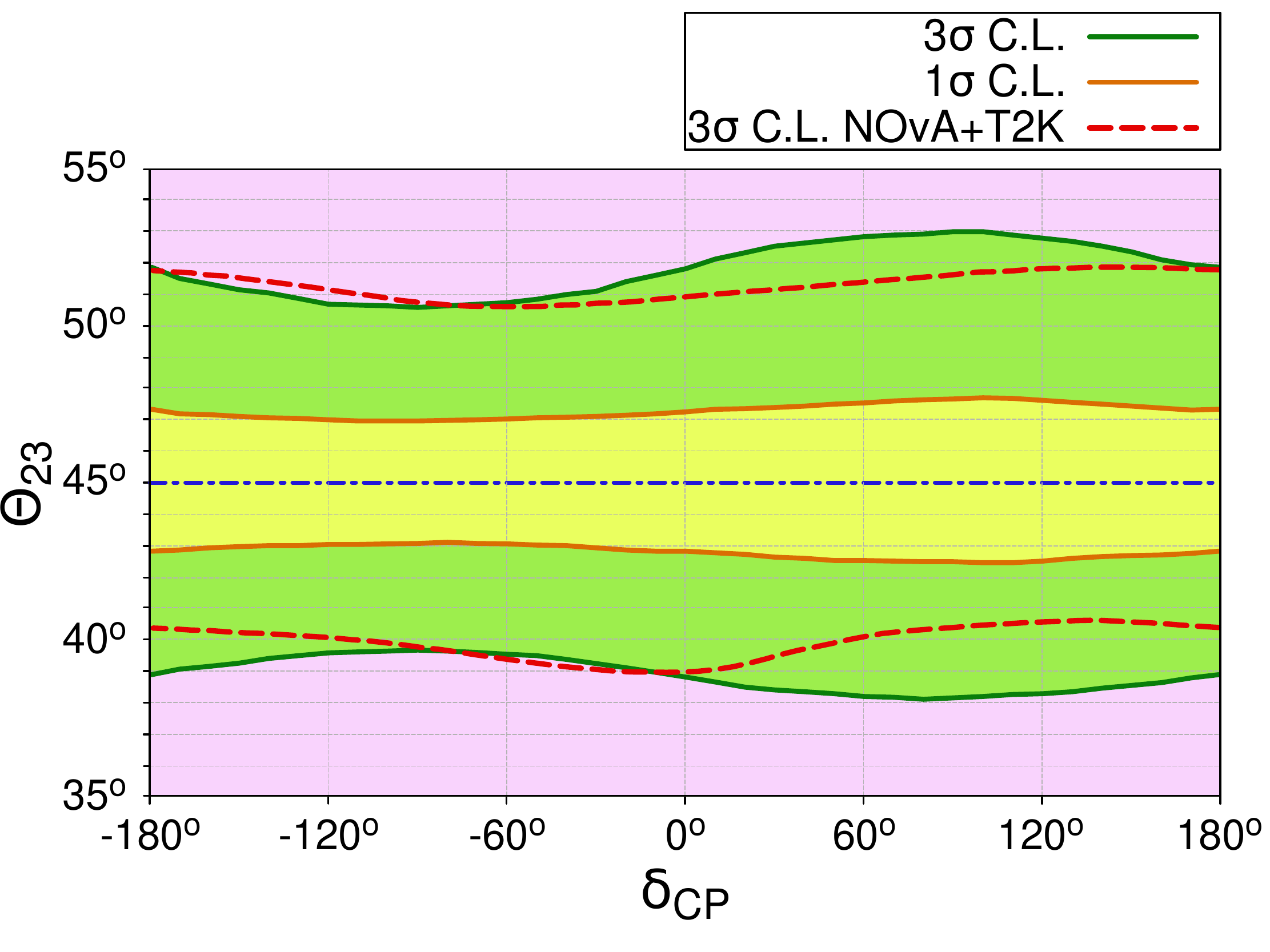} 
\end{minipage}
\caption{\label{fig:1}The sensitive area for the determination of octant as a function of $\delta_{\rm CP}$ assuming an exposure time of 200 kt$\cdot$year. The left panel corresponds to 5 MW source power, the right panel -- 10 MW. 
The yellow area corresponds to insensitivity with 68.3\% C.L. The green area corresponds to insensitivity with 99.7\% C.L. The pink area is sensitive to the octant with significance more than 3$\sigma$. Dashed red lines show 99.7\% C.L for combined analysis of T2K and NO$\nu$A, which is presented in~\cite{Agarwalla}.}
\end{figure}

Experimental sensitivity to octant degeneracy is depicted in Figure~\ref{fig:1}. 
The yellow area shows the 68.3\% confidence interval,  within which the experiment is insensitive to octant degeneracy.
The green area shows the insensitive region with confidence level 99.7\%. 
In the pink area the octant can be determined with significance more than 3$\sigma$.
As can be seen the sensitivity to octant is better for negative values of  $\delta_{\rm CP}$.
For these values a 5 MW cyclotron can distinguish the octant, if the mixing angle $\Theta_{23}$ is outside the range $38.5^{\rm o}-52.9^{\rm o}$.  	
Whereas a 10 MW cyclotron can measure the octant if $\Theta_{23}$  is outside $39.7^{\rm o}-50.8^{\rm o}$. 
Therefore increasing statistics causes an improvement in the sensitivity. 
The result for the 10 MW case is slightly worse than the expected result from the combined analysis of T2K+NO$\nu$A.

\begin{figure}[ht]
\centering
\begin{minipage}[b]{0.44\linewidth}
\centering
\includegraphics[scale=0.395]{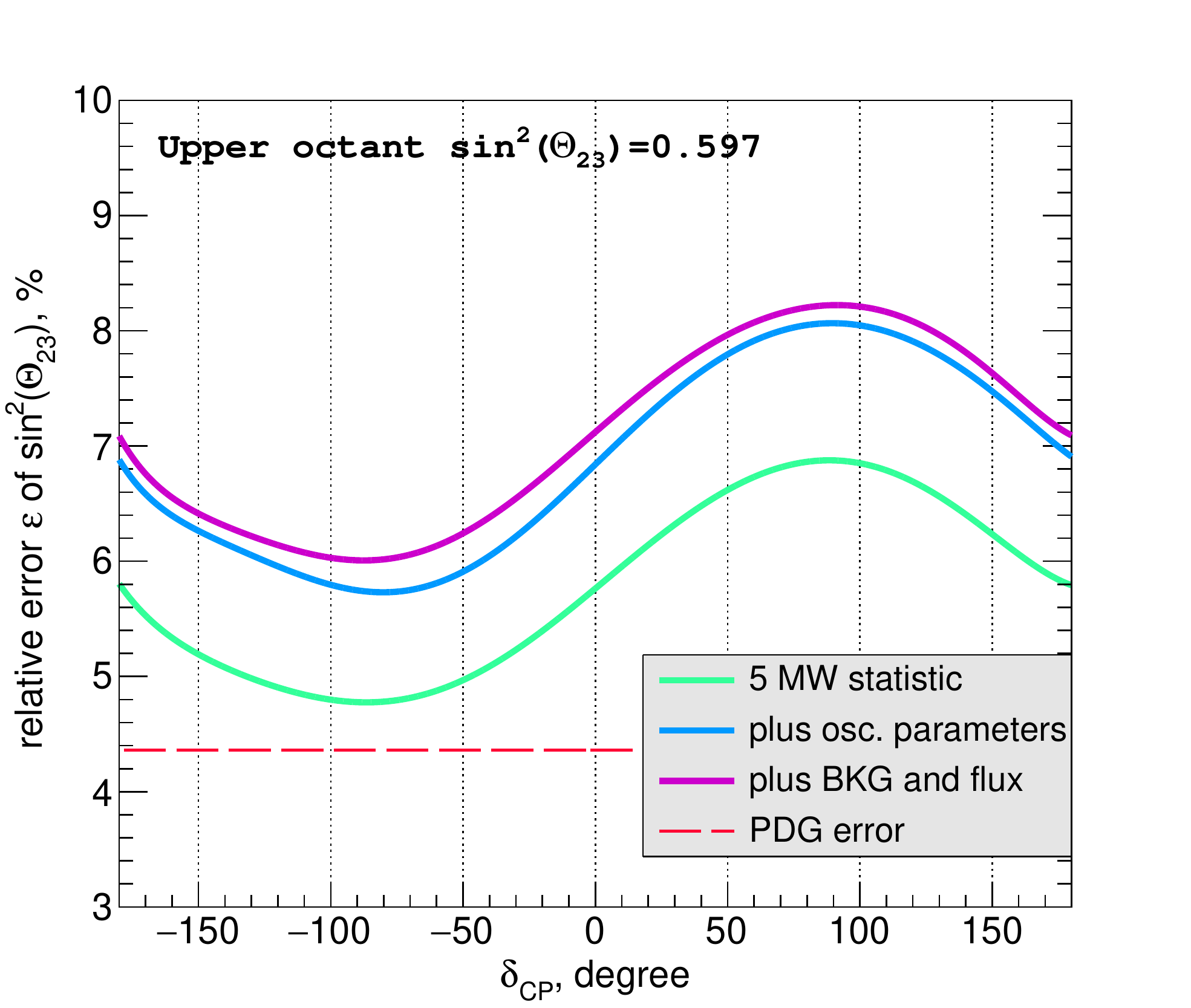}
\end{minipage}
\qquad
\begin{minipage}[b]{0.489\linewidth}
\centering
\includegraphics[scale=0.395]{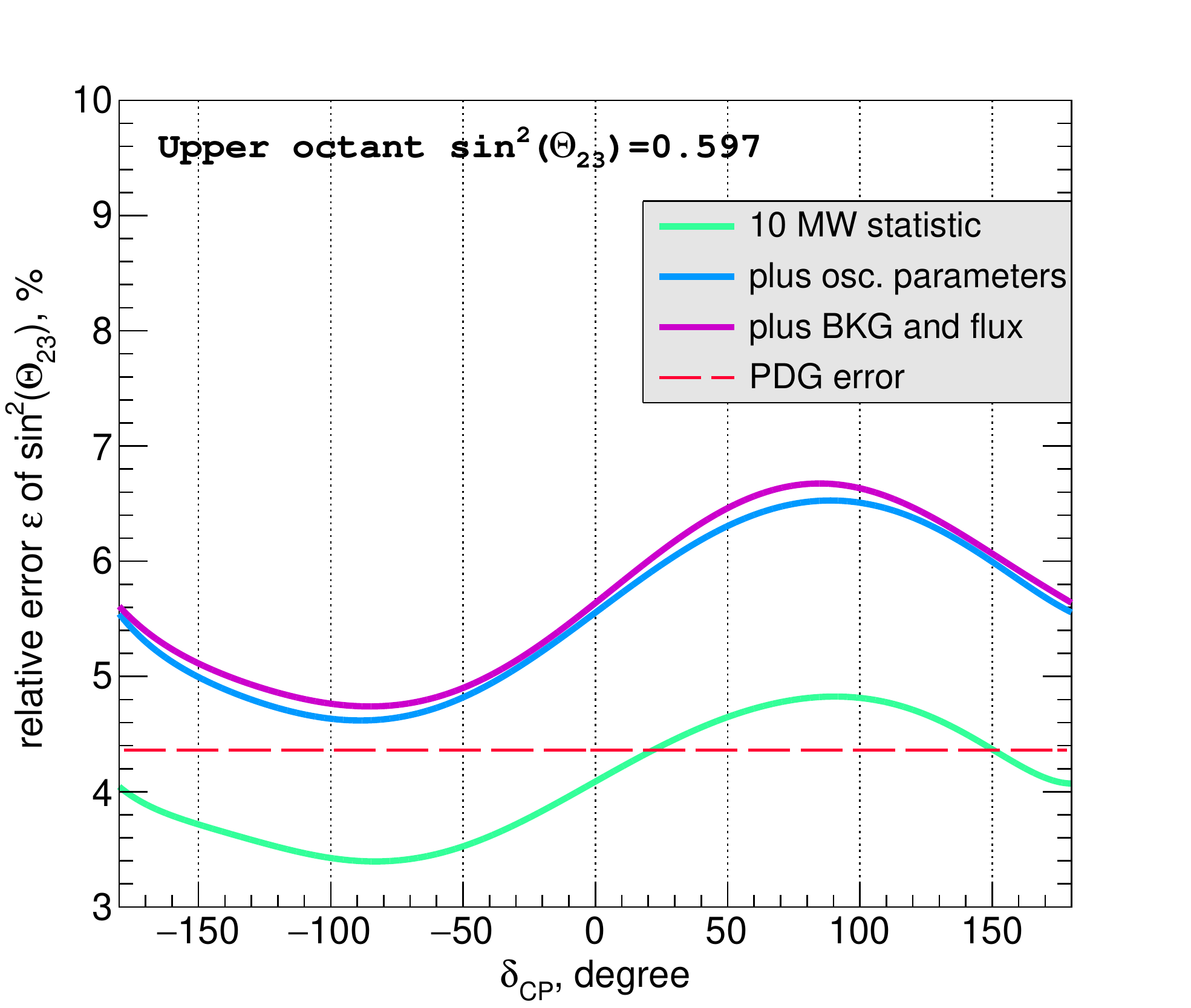} 
\end{minipage}
\vfill
\begin{minipage}[b]{0.44\linewidth}
\centering
\includegraphics[scale=0.395]{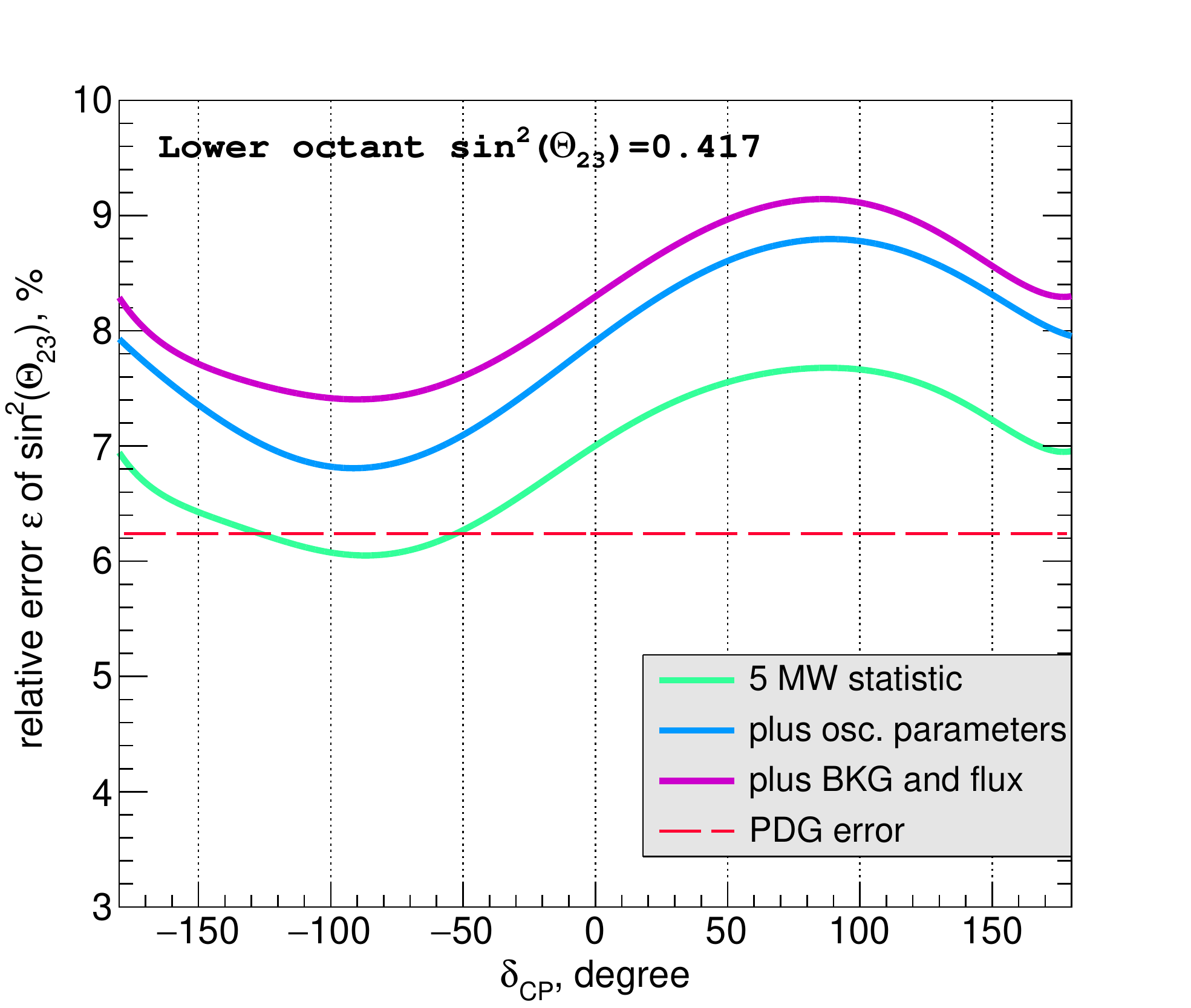}
\end{minipage}
\qquad
\begin{minipage}[b]{0.489\linewidth}
\centering
\includegraphics[scale=0.395]{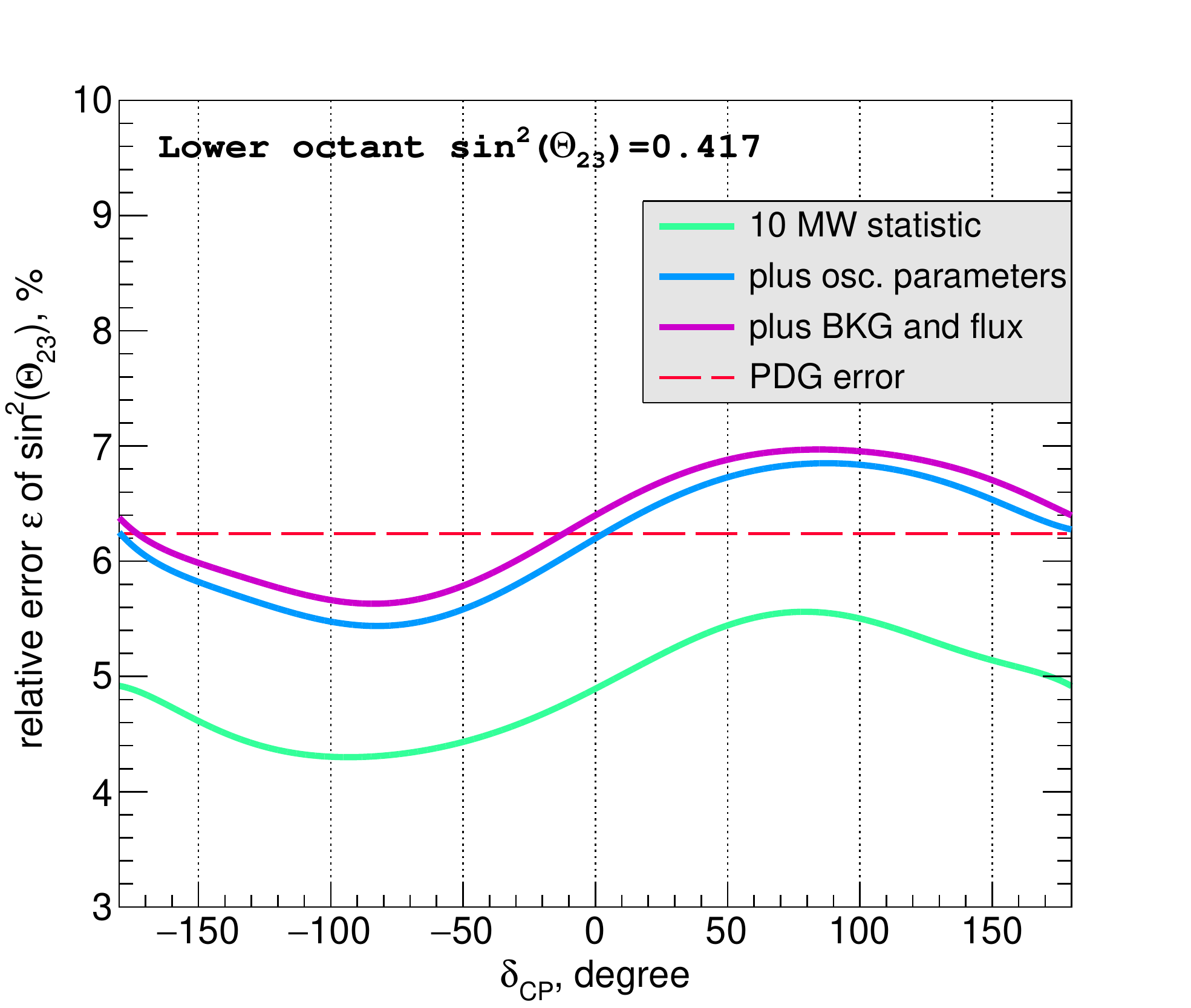} 
\end{minipage}
\caption{\label{fig:2}Accuracy of measuring mixing angle $\Theta_{23}$. The top row is for $\sin^2(\Theta_{23})= 0.597$ and the bottom for $\sin^2(\Theta_{23})= 0.417$. Two powers of far cyclotron are assumed. 
Dashed white lines corresponds to the current value of the relative error $\sin^2(\Theta_{23})$ from PDG, where $\varepsilon(\sin^2(\Theta_{23})= 0.417)=6.24\%$ and $\varepsilon(\sin^2(\Theta_{23})= 0.597)=4.36\%$.}
\end{figure}

Figure~\ref{fig:2} gives the quantitative estimation of the uncertainty for two possible values of $\Theta_{23}$ as a function of $\delta_{\rm CP}$.  
The top row corresponds to $\sin^2(\Theta_{23})= 0.597$, and the bottom row to $\sin^2(\Theta_{23})= 0.417$.
The wave behavior of curves in Fig.~\ref{fig:2} can be explained by maximum of the probability function~\eqref{eq_21} for $\delta_{\rm CP}=\pi/2$ and minimum for $\delta_{\rm CP}=-\pi/2$.
As can be seen in Figure \ref{fig:2}, the main uncertainty comes from oscillation parameters.
Our estimation shows that the dominant uncertainty comes from mixing angle $\Theta_{13}$.
The influence of background is quite small, especially for higher statistics with a 10 MW source.
Statistically the improvement in the result is possible only for the 10 MW source.
However, in reality only values from the LO can improve the result in the case of negative CP-phase.

\section{Conclusion}
\par
\indent

Current work has demonstrated another application of using a superconductive cyclotron for measurements in neutrino physics. 
The transition channel  ${\bar{\nu }}_{\mu }{\to }{\bar{\nu }}_{e}$ allows us to explore not only the problem of CP violation, but at the same time to realize the precise measurement of $\Theta_{23}$ and partially resolve octant degeneracy. 

It was shown that the distinction between LO and UO is comparable with combined analysis of T2K and NO$\nu$A especially for negative values of  $\delta_{\rm CP}$.
Regarding the precision of $\Theta_{23}$ measuring, the current best fit value can be improved only for the 10 MW case, especially if the mixing angle lies in LO.
There are two main difficulties for precision measurements: uncertainties in oscillation parameters and small statistics.  
The problem of statistics can be reduced through the use of a small water detector instead of the near cyclotron for monitoring neutrino flux. This allows us to use the far cyclotron continuously, as proposed for the TNT2K experiment~\cite{tnt2k}. 

The combination of JUNO and superconductive cyclotrons can be a good alternative to conventional beam experiments. It will allow for the measurement of $\Theta_{23}$ and $\delta_{\rm CP}$ in the current scientific program without the spoiling of JUNO's main goals.

\acknowledgments

This work was in part supported by 
the China Postdoctoral Science Foundation under Grant No. 2018M643283, 
the National Key R\&D program of China under Grant No. 2018YFA0404103 and the National Natural Science Foundation of China (NSFC) under Grant No. 11775315. 
We would like to extend great thanks to School of Physics, Sun Yat-Sen University,
especially to the leader of our neutrino group Prof. Wei Wang for cultivating good working conditions.
We also express special gratitude to Dr. Neill Raper for editing of this paper.


\begin{thebibliography}{99}

\bibitem{PMNS}
B. Pontecorvo, 
\emph{Zh. Eksp. Teor. Fiz.} {\bf 33} (1957) 549, [JETP {\bf6}, 429 (1958)]; 
Z. Maki, M. Nakagawa, and S. Sakata, 
\emph{Prog. Theor. Phys.} {\bf 28} (1962) 870. 

\bibitem{MINOS_2014}
P.~Adamson {\it et al.} [MINOS Collaboration],
\emph{Combined analysis of $\nu_{\mu}$ disappearance and $\nu_{\mu} \rightarrow \nu_{e}$ appearance in MINOS using accelerator and atmospheric neutrinos},
\emph{Phys.\ Rev.\ Lett.}  {\bf 112} (2014) 191801,
arXiv:1403.0867.

\bibitem{T2K}
K. Abe {\it et al.} [T2K Collaboration],
\emph{Combined analysis of neutrino and antineutrino oscillations at T2K}, 
\emph{Phys. Rev. Lett.} {\bf118} (2017) 151801.



\bibitem{NOvA_2018}
M.A. Acero {\it et al.} [NOvA Collaboration],
\emph{New constraints on oscillation parameters from $\nu_e$ appearance and $\nu_\mu$ disappearance in the NOvA experiment},
\emph{Phys. Rev. D} {\bf 98} (2018) 032012,
arXiv:1806.00096.

\bibitem{juno}
Fengpeng An {\it et al.}, \emph{Neutrino physics with JUNO}, 
\emph{J. Phys. G: Nucl. Part. Phys.} {\bf43}, (2016) 030401.

\bibitem{CPV_my}
M.V. Smirnov, Z.J. Hu, S.J. Li and J.J. Ling,
\emph{The possibility of leptonic CP-violation measurement with JUNO},
\emph{Nucl.\ Phys.\ B} {\bf 931}, (2018) 437,
arXiv:1802.03677.

\bibitem{daedalus} 
J.~Alonso {\it et al.},
\emph{Expression of Interest for a Novel Search for CP Violation in the Neutrino Sector: DAE$\delta$ALUS}, (2010),
arXiv:1006.0260.

\bibitem{Bilenky}
S.M. Bilenky,
\emph{On the phenomenology of neutrino oscillations in vacuum},
(2012), arXiv:1208.2497.

\bibitem{conrad}
J.M. Conrad and M.H. Shaevitz,
\emph{Multiple Cyclotron Method to Search for CP Violation in the Neutrino Sector},
\emph{Phys. Rev. Lett.} {\bf 104}, (2010) 141802.

\bibitem{wilks}
S.S. Wilks, 
\emph{The Annals of Mathematical Statistics}, {\bf9}, (1938).

\bibitem{num_stat}
W.H. Press, S.A. Teukolsky, W.T. Vetterling, B.P. Flannery, 
\emph{Numerical Recipes in C: The Art of Scientific Computing}, ISBN 0-521-43108-5,
(1992).

\bibitem{Das_lbno}
C.R. Das, J. Maalampi, J. Pulido and S. Vihonen,
\emph{Determination of the $\Theta_{23}$ octant in LBNO},
\emph{JHEP} {\bf 1502} (2015) 048,
arXiv:1411.2829.

\bibitem{Huber}
P. Huber, J. Kopp, M. Lindner, M. Rolinec and W. Winter,
\emph{New features in the simulation of neutrino oscillation experiments with GLoBES 3.0: General Long Baseline Experiment Simulator},
\emph{Comput. Phys. Commun.}  {\bf 177} (2007) 432,
arXiv:hep-ph/0701187.

\bibitem{PDG} 
M. Tanabashi {\it et al.} [Particle Data Group], 
\emph{Phys. Rev. D} {\bf98}, (2018) 030001.

\bibitem{Randolph}
R. Mollenberg, F. von Feilitzsch, D. Hellgartner, L. Oberauer, M. Tippmann, V. Zimmer, J. Winter and M. Wurm,
\emph{Detecting the Diffuse Supernova Neutrino Background with LENA},
\emph{Phys.\ Rev.\ D} {\bf 91} (2015) no.3,  032005.
arXiv:1409.2240.

\bibitem{minuit}
F.~James and M.~Roos,
\emph{Minuit: A System for Function Minimization and Analysis of the Parameter Errors and Correlations},
\emph{Comput. Phys. Commun.}  {\bf 10}, (1975) 343.

\bibitem{minuit2}
F.~James and M.~Winkler,
\emph{MINUIT User's Guide}, (2004).

\bibitem{Agarwalla}
S.K. Agarwalla, S. Prakash and S.U. Sankar,
\emph{Resolving the octant of $\theta_{23}$ with T2K and NOvA},
\emph{JHEP} {\bf 1307} (2013) 131,
arXiv:1301.2574.

\bibitem{tnt2k}
J. Evslin, S.F. Ge and K. Hagiwara,
\emph{The leptonic CP phase from T2(H)K and $\mu^{+}$ decay at rest},
\emph{JHEP} {\bf 1602} (2016) 137,
arXiv:1506.05023.

%
%
%
%
%
%
%
%
%
%
%
%
%





\end{thebibliography}
\end{document}